\documentclass[showpacs,10pt,twocolumn,prb]{revtex4-1}
\usepackage{amsmath}
\usepackage{amssymb}
\usepackage{graphics}
\usepackage{epsfig}

\begin{document}

\title{Thermally activated energy and flux flow Hall effect of Fe$_{1+y}$(Te$%
_{1-x}$S$_{x}$)$_{z}$}
\author{Hechang Lei,$^{1}$ Rongwei Hu,$^{1,\ast }$ E. S. Choi,$^{2}$ and C.
Petrovic$^{1}$}
\date{\today}

\begin{abstract}
Thermally activated flux flow (TAFF) and flux flow Hall effect (FFHE) of
Fe(Te,S) single crystal in the mixed state are studied in magnetic fields up
to 35 T. Thermally activated energy (TAE) is analyzed using conventional
Arrhenius relation and modified TAFF theory which is closer to experimental
results. The results indicate that there is a crossover from single-vortex
pinning region to collective creep pinning region with increasing magnetic
field. The temperature dependence of TAE is different for $H//ab$ and $%
H//c$. On the other hand, the analysis of FFHE in the mixed state indicates
that there is no Hall sign reversal. We also observe scaling behavior $|\rho
_{xy}(H)|=A\rho _{xx}(H)^{\beta }$.
\end{abstract}


\affiliation{$^{1}$Condensed Matter Physics and Materials Science Department, Brookhaven
National Laboratory, Upton, NY 11973, USA}
\affiliation{$^{2}$NHMFL/Physics, Florida State University, Tallahassee, Florida 32310,
USA}

\maketitle

\section{Introduction}

The discovery of iron based materials has generated enormous interests in
the field of superconductivity.\cite{Kamihara}$^{-}$\cite{Hsu FC} Due to
similar layered structure to cuprate oxides and rather high $T_{c}$, iron
based superconductors could host rich vortex phenomena in the mixed state.%
\cite{Blatter}$^{-}$\cite{Fisher} Recent work suggests that vortex
properties of iron based materials seem to be similar to the cuprate
superconductors since magnetic flux collective pinning and creep region as
well as fishtail effects (second peak effect) have been observed in LnOFeAs
(Ln=rare earth elements, 1111 system) and AFe$_{2}$As$_{2}$ (A=alkaline
earth elements, 122 system).\cite{Yang H}$^{-}$\cite{Jaroszynski}

Iron based superconductors, FeSe$_{1-x}$, Fe$_{1+y}$Te$_{1-x}$Se$_{x}$, and
Fe$_{1+y}$Te$_{1-x}$S$_{x}$ (11 system),\cite{Hsu FC}$^{,}$\cite{Yeh KW}$%
^{-} $\cite{McQueen} are of interest both for the technological applications
and for the understanding of the vortex properties in the mixed state due to
rather simple structure and nearly isotropic upper critical field.\cite{Fang}%
$^{-}$\cite{Lei HC} Only a limited amount of information on the vortex
behavior in single crystals of 11 system is available until now, mainly
focusing on thermally activated flux flow (TAFF) region in Fe(Te,Se).\cite%
{Yadav}$^{,}$\cite{Yadav2} Normal carriers in the vortex core, which
experience a Lorentz force, can lead to normal Hall effect in mixed state.
On the other hand, flux flow can also induce Hall effect in the mixed state.
In detail, when applying a transport current, the flux lines will experience
the Lorentz force, $\mathbf{F}=\frac{1}{c}\mathbf{j}\times \mathbf{B}$,
where $\mathbf{j}$ is the supercurrent density and $\mathbf{B}$ is the
magnitude of magnetic induction. The motion of magnetic flux lines produces
a macroscopic electric field $\mathbf{E}$ which is given by $\mathbf{E}=-%
\frac{1}{c}\mathbf{v}\times \mathbf{B}$, where $\mathbf{v}$ is the velocity
of vortex motion.\cite{Josephson} The vortex motion along the Lorentz force
(perpendicular to $\mathbf{j}$) gives the dissipative field ($\mathbf{E\Vert
j}$) and generates the flux flow resistivity, whereas the vortex motion
along the direction of supercurrent results in the Hall electric field ($%
\mathbf{E\bot j}$). Therefore the flux flow Hall effect (FFHE) is a
sensitive method to study the vortex dynamics.

There are two exotic phenomena in connection with FFHE in cuprate
superconductors. One is a sign reversal of the Hall resistivity $\rho
_{xy}(H)$ below $T_{c}$.\cite{Hagen}\ This anomaly has also been observed in
some conventional superconductors, e.g. amorphous MoSi$_{3}$,\cite{Smith}
and 2H-NbSe$_{2}$.\cite{Bhattacharya} The sign change is not expected within
the classical Bardeen-Stephen\cite{Bardeen} and Nozi\`{e}res-Vinen\cite%
{Nozieres} theories of vortex motion, which predict that the Hall sign in
the superconducting and normal state should be the same. Several models have
been proposed for interpreting this anomaly,\cite{Dorsey}$^{-}$\cite{Otterlo}
however its origin remains a controversy. Another phenomenon is a scaling
law between $\rho _{xy}(H)$ and the longitudinal resistivity $\rho _{xx}(H)$
in the superconducting transition region, i.e., $|\rho _{xy}(H)|=A\rho
_{xx}(H)^{\beta }$ with different values of $\beta $ for different materials.%
\cite{Luo}$^{,}$\cite{Samoilov}

In this paper, we study the vortex properties of Fe$_{1.14(1)}$(Te$%
_{0.91(2)} $S$_{0.09(2)}$)$_{z}$ single crystal via TAFF resistivity and
FFHE in the mixed state. The temperature dependence of TAE is different for $%
H\Vert ab$ and $H\Vert c$. Furthermore, there is a crossover from
single-vortex pinning region to collective creep pinning region with
increasing magnetic field. On the other hand, there is no sign reveral and
we observe scaling behavior for FFHE.

\section{Experiment}

Single crystals of Fe(Te,S) were grown by self flux method and their crystal
structure was analyzed in the previous report.\cite{Hu RW} The elemental
analysis of the crystal used in this study showed the stoichiometry is
Fe:Te:S=1.14(1):0.91(1):0.09(2) and we denote it as S-09 in the following
for brevity. Electrical transport measurements were performed using a
four-probe configuration with current flowing in the ab-plane of tetragonal
structure in dc magnetic fields up to 9 T in a Quantum Design PPMS-9 from
1.9 to 200 K and up to 35 T in a He3 cryostat system with resistive magnet
down to 0.3 K at the National High Magnetic Field Laboratory (NHMFL) in
Tallahassee, FL. Hall contacts with typical misalignment of less than 0.1 mm
were used. At each point the Hall voltage was measured for two directions of
the magnetic field, which is always perpendicular to current direction.

\section{Results and Discussions}

Fig. 1(a,b) show the resistivity $\rho (T,H)$ of S-09 near the
superconducting transition region for $H\parallel ab$ plane and $H\parallel
c $ axis. With increasing magnetic fields, the resistivity transition widths
are broadened gradually. The onset of superconductivity shifts to lower
temperatures for both magnetic field directions, but the trend is more
obvious for $H\Vert c$ than $H\Vert ab$. According to the TAFF theory, the
resistivity in TAFF region can be expressed as,\cite{Blatter}$^{,}$\cite%
{Palstra1}$^{,}$\cite{Palstra2}%
\begin{equation}
\rho =(2\nu _{0}LB/J)exp(-J_{c0}BVL/T)sinh(JBVL/T)
\end{equation}

where $\nu _{0}$ is an attempt frequency for a flux bundle hopping, $L$ is
the hopping distance, $B$ is the magnetic induction, $J$ is the applied
current density, $J_{c0}$ is the critical current density in the absence of
flux creep, $V$ is the bundle volume and $T$ is the temperature. If $J$ is
small enough and $JBVL/T\ll 1$, we obtain%
\begin{equation}
\rho =(2\rho _{c}U/T)exp(-U/T)=\rho _{0f}exp(-U/T)
\end{equation}

where $U=J_{c0}BVL$ is the thermally activated energy (TAE) and $\rho
_{c}=\nu _{0}LB/J_{c0}$, which is usually considered to be
temperature-independent. For cuprate superconductors, the prefactor $2\rho
_{c}U/T$ is usually assumed as a constant $\rho _{0f}$,\cite{Palstra1}
therefore, $ln\rho (T,H)=ln\rho _{0f}-U(T,H)/T$, where $H$ is the external
magnetic field. On the other hand, according to the condensation model,\cite%
{Palstra2} $U(T,H)=H_{c}^{2}(t)\xi ^{n}(t)$, where $H_{c}$ is the thermal
critical field, $\xi $ is the coherence length, $t=T/T_{c}$ ($T_{c}$ is the
superconducting transition temperature), and $n$ depends on the
dimensionality of the vortex system with the range from 0 to 3. Since $%
H_{c}\propto 1-t$, and $\xi \propto (1-t)^{-1/2}$ near $T_{c}$,\cite{Brandt}
it is obtained that $U(T,H)=U_{0}(H)(1-t)^{q}$, where $q=2-n/2$. It is
generally assumed that $U(T,H)=U_{0}(H)(1-t)$, i.e. $n=2$, and the $ln\rho
-1/T$ becomes Arrhenius relation, $ln\rho (T,H)=ln\rho _{0}(H)-U_{0}(H)/T$,
where $ln\rho _{0}(H)=ln\rho _{0f}+U_{0}(H)/T_{c}$ and $U_{0}(H)$ is the
apparent activation energy. Furthermore, it can be concluded that $-\partial
ln\rho (T,H)/\partial T^{-1}=U_{0}(H)$. Hence, the $ln\rho$ vs. $1/T$ should
be linear in TAFF region. The slope is $U_{0}(H)$ and its y-intercept
represents $ln\rho _{0}(H)$.

\begin{figure}[tbp]
\centerline{\includegraphics[scale=0.95]{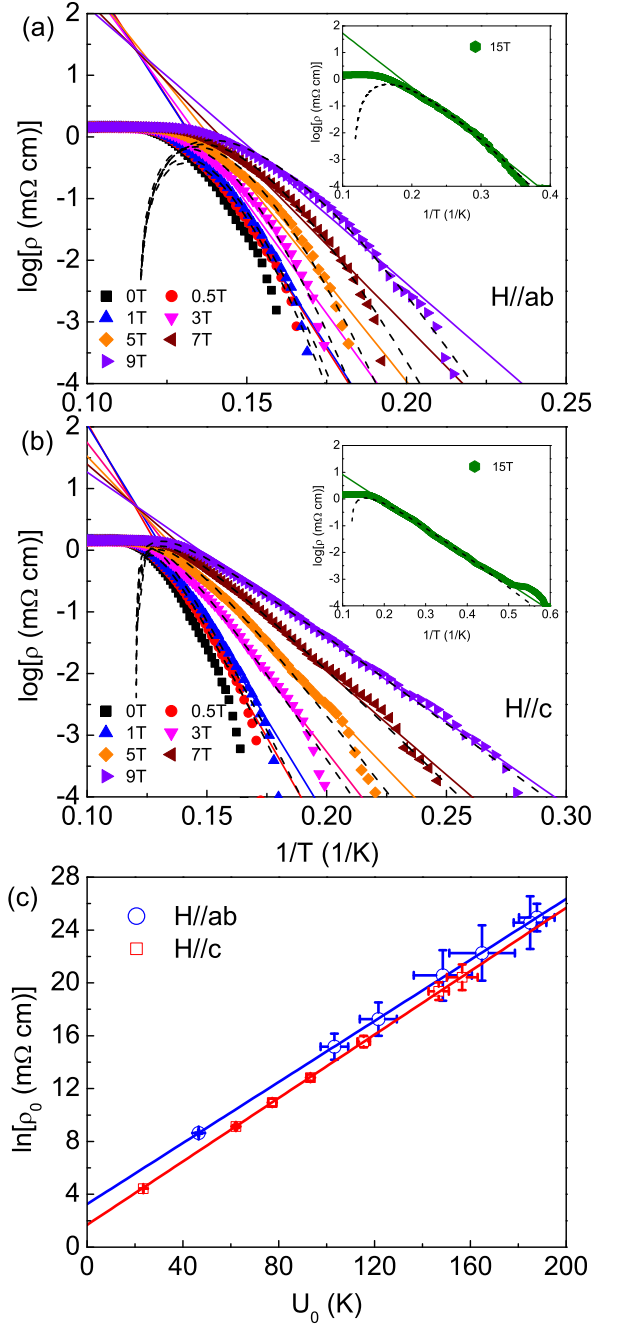}} \vspace*{-0.3cm}
\caption{(a) and (b) Longitudinal resistivities $\protect\rho (T,H)$ of S-09
in different magnetic field directions for $H\Vert ab$ and $H\Vert c$ below
10K, respectively. The corresponding solid and black dashed lines are
fitting results from the Arrhenius relation\ and eq. (3). (c) $ln\protect%
\rho _{0}$ vs $U_{0}$ derived from Arrhenius relation for $H\Vert ab$ and $%
H\Vert c$.}
\end{figure}

In the Figs. 1(a,b), the solid lines show the Arrhenius relation in TAFF
region. Note that the results are shown in the common logarithmic scale in
the figures, but we calculate them in the natural one. All linear fits
intersect at approximately the same point $T_{cross}$, which is about $8.63$
and $8.34$ K for $H\parallel ab$ and $H\parallel c$, respectively. Assuming
the temperature dependences of $\rho _{ab}(T,H)$ at two different magnetic
fields ($H_{1}$ and $H_{2}$) can be fitted by Arrhenius relation, according
to above discussion, we get $ln\rho (T,H_{1})=ln\rho
_{0f}+U_{0}(H_{1})/T_{c}-U_{0}(H_{1})/T$ and $ln\rho (T,H_{2})=ln\rho
_{0f}+U_{0}(H_{2})/T_{c}-U_{0}(H_{2})/T$. When $\rho (T,H_{1})=\rho (T,H_{2})
$, it can be obtained $T=T_{c}$, therefore, ideally, all the lines at
different fields should be crossed into same point, $T_{cross}$, which is
equal to $T_{c}$. According to the conventional analysis, the $ln\rho _{0f}$
and $T_{c}$ can be obtained from linear fits $ln\rho _{0}(H)$ and $U_{0}(H)$
using $ln\rho _{0}(H)=ln\rho _{0f}+U_{0}(H)/T_{c}$ (shown in Fig. 1(c)).
From the fitting results, values of $\rho _{0f}$ and $T_{c}$ are $27.22\pm
4.77$ $m\Omega \cdot cm$, $8.72\pm 0.82$ K and $5.23\pm 1.42$ $m\Omega \cdot
cm$, $8.32\pm 0.48$ K for $H\parallel ab$ and $H\parallel c$, respectively.
The $T_{c}$ is consistent with the values of $T_{cross}$ in the range of
errors. It seems that the $\rho (T,H)$ can be fitted with straight lines
well. However, close inspection shows that there is rather large fitting
errors, especially for $H\parallel ab$. The origin of the large errors is
the Arrhenius relation that can only be satisfied in the limited region and this
region is narrower for $H\parallel ab$. The effects of prefactor and
non-linear relation of $U(T,H)$ lead to $\rho (T,H)$ deviating from
Arrhenius relation (vide infra).

Fig. 2(a,b) shows the temperature dependence of $-\partial ln\rho
(T,H)/\partial T^{-1}$ for both field directions. Because the assumptions $%
U(T,H)=U_{0}(H)(1-t)$ and $\rho _{0f}=const$ lead to $-\partial ln\rho
(T,H)/\partial T^{-1}=U_{0}(H)$, $U_{0}(H)$ should be a set of horizontal
lines. We present this in Fig. 2(a,b) over a limited length. Each length
covers the temperature interval used for estimating $U_{0}(H)$ in the
Arrhenius relation. It can be seen that $-\partial ln\rho (T,H)/\partial
T^{-1}$ increases sharply with decreasing temperature, which was also
observed in Bi-2212 thin films.\cite{Zhang YZ1} The center of each $U_{0}(H)$
horizontal line approximately intersects $-\partial ln\rho (T,H)/\partial
T^{-1}$ curve and the overlapping region is increasing with temperature
decrease. This shows that each $U_{0}(H)$ is only the average value of its $%
-\partial ln\rho (T,H)/\partial T^{-1}$ in the fitting temperature region.
Hence, the TAE determined from the conventional method does not reflect the
true evolution of $U(T,H)$ with the temperature, particularly for $%
H\parallel ab$. This contradiction originates from two basic assumptions
introduced for Arrhenius relation: one is the constant prefactor $\rho
_{0f}=2\rho _{c}U/T$ and another is the linear relation $U(T,H)=U_{0}(H)(1-t)
$. Zhang \textit{et al}\cite{Zhang YZ2} suggested that the
temperature-dependent prefactor and nonlinear relation of $U(T,H)-T$ should
be considered. In the following section, we will analyze the resistivity
results using this more general method.\cite{Zhang YZ2}

Using the relation $U(T,H)=U_{0}(H)(1-t)^{q}$, from eq. (2) it can be
derived that

\begin{equation}
ln\rho =ln(2\rho _{c}U_{0})+q\ln (1-t)-\ln T-U_{0}(1-t)^{q}/T
\end{equation}

and%
\begin{equation}
-\partial ln\rho /\partial T^{-1}=[U_{0}(1-t)^{q}-T][1+qt/(1-t)]
\end{equation}

where $\rho _{c}$ and $U_{0}$ are temperature independent and $T_{c}$
derived from Arrhenius relation is used for fitting. Therefore, there are
three free parameters, $q$, $\rho _{c}$ and $U_{0}$ in eq. (3). The fitting
results are shown in Fig. 1. From Fig. 1(a,b), it can be seen that all fits
are in good agreement with experimental data and the results are better than
Arrhenius relation. This is more pronounced for $H\parallel ab$. Fig. 2(a,b)
clearly shows the advantage of eq. (3) over Arrhenius relation. The TAFF
formula (eq. (1)) can effectively capture the upturn trend of $-\partial
ln\rho /\partial T^{-1}$ with decreasing temperature when the (linear or
nonlinear) correlations between prefactor as well as $U(T,H)$ and $T$ are
considered. In detail, when $T\ll U$ (corresponding to $T\ll U_{0}(1-t)^{q})$%
, it can be derived that $-\partial ln\rho /\partial
T^{-1}=U_{0}(1-t)^{q}[1+qt/(1-t)]$ and when $q=1$, $-\partial ln\rho
/\partial T^{-1}=U_{0}$, i.e. Arrhenius relation. Because the obtained $%
U_{0} $ of S-09 is much smaller than that of cuprates superconductors (shown
in Fig. 3) and it is comparable with temperature, the assumption $T\ll U$
can not be satisfied and the temperature dependence of prefactor should be
considered.

\begin{figure}[tbp]
\centerline{\includegraphics[scale=0.95]{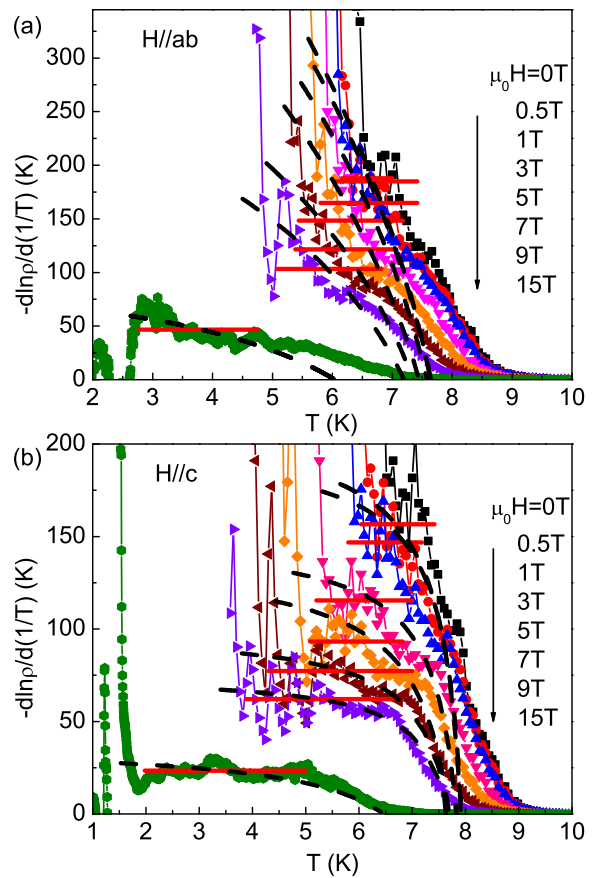}} \vspace*{-0.3cm}
\caption{Experimental $-\partial ln\protect\rho /\partial T^{-1}$ data in
TAFF region for $H\Vert ab $ (a) and $H\Vert c$ (b), respectively. The red
solid horizontal lines correspond to obtained $U_{0}(H) $ from Arrhenius
relation and the blue dashed lines are plotted using eq. (4). The parameters
are determined via fitting eq. (3) to corresponding experimental resistivity
data shown in Fig.1 (a) and (b).}
\end{figure}

Fig. 3 presents the $U_{0}(H)$ and $q(H)$ obtained from experimental data
fits using eq. (3) at different fields. The $U_{0}(H)$ shows a power law ($%
U_{0}(H)\sim H^{-\alpha }$) field dependence for both directions. For $%
H\Vert ab$, $\alpha =0.12\pm 0.02$ for $\mu _{0}H<5T$ and $\alpha =1.70\pm
0.30$ for $\mu _{0}H>5T$; For $H\Vert c$, $\alpha =0.21\pm 0.03$ for $\mu
_{0}H<5T$ and $\alpha =1.34\pm 0.16$ for $\mu _{0}H>5T$. The weak field
dependence of $U_{0}(H)$ in low field for both orientations suggests that
single-vortex pinning dominates in this region.\cite{Blatter} The vortex
spacing becomes significantly smaller than penetration depth in higher
fields and we expect a crossover to a collective-pinning regime where the
activation energy becomes strongly dependent on the field, i.e., the
collective creep dominance.\cite{Yeshurun} Because $\alpha $ is larger than $%
1$ at this regime, it is possible that the flux lines are pinned by the
collective point defects in the high field region.\cite{Chin} Similar
crossover has been observed in Nd(O,F)FeAs single crystal.\cite{Jaroszynski2}
The values of $q$ change from about $1$ for $H\Vert c$ to $2$ for $H\Vert ab$%
, independent on the intensity of field for both directions. The value of $%
q=2$ has also been observed in many cuprates superconductors.\cite{Zhang YZ1}%
$^{,}$\cite{Zhang YZ2}$^{,}$\cite{Wang}

\begin{figure}[tbp]
\centerline{\includegraphics[scale=0.8]{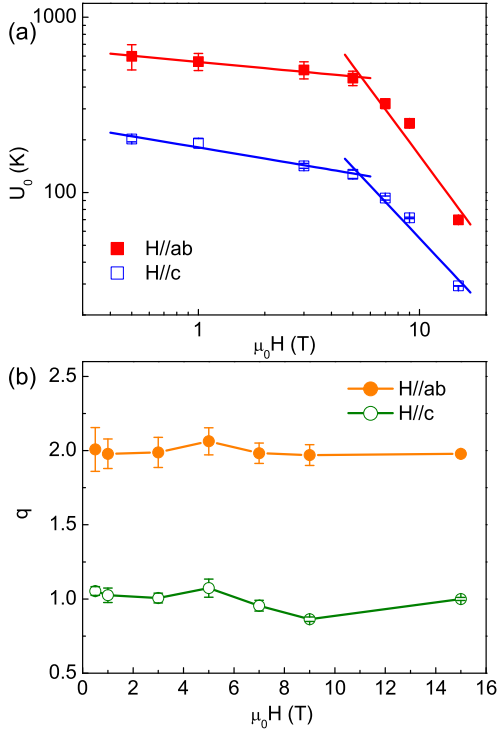}} \vspace*{-0.3cm}
\caption{(a) $U_{0}$ and (b) $q$ as functions of magnetic fields obtained
from fitting the resistivity in TAFF region using eq. (3). The opened and
filled squares represent $U_{0}$ for $H\Vert c$ and $H\Vert ab$,
respectively, while the opened and filled circles show corresponding q,
respectively. The solid lines in (a) are power-law fitting using $%
U_{0}(H)\sim H^{-\protect\alpha}$.}
\end{figure}

\begin{figure}[tbp]
\centerline{\includegraphics[scale=0.95]{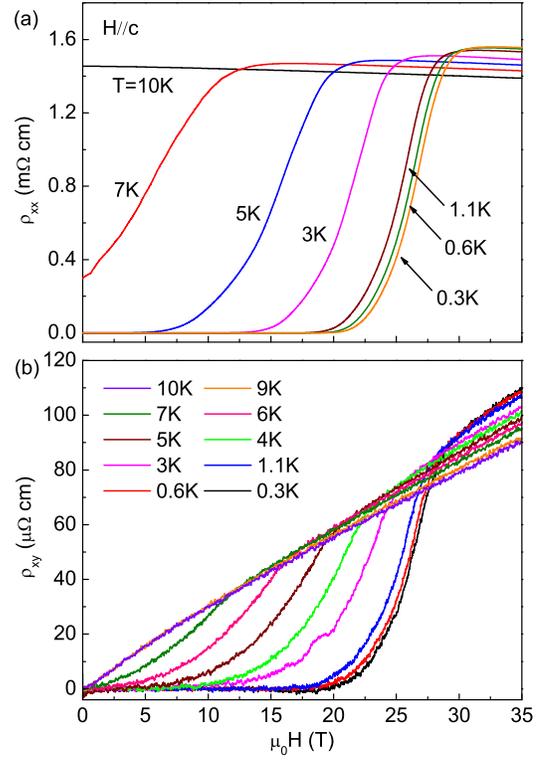}} \vspace*{-0.3cm}
\caption{Field dependence of (a) Longitudinal resistivity $\protect\rho %
_{xx}(H)$ and (b) Hall resistivity $\protect\rho _{xy}(H)$ at various
temperatures in dc magnetic fields up to 35T for $H\Vert c$.}
\end{figure}

It should be noted that although the Fe(Te,S) sample has low volume
fraction, it should not have obvious effect on the analysis of TAFF. Due to
the inhomogeneity of sample, the conductivity of sample can be expressed as $%
\sigma =\sigma _{sc}+\sigma _{normal}$, where $\sigma _{sc}$ is the
conductivity in superconducting state and $\sigma _{normal}$ is the
conductivity of normal state. When $T<T_{c,zero}$, $\sigma _{sc}$ is
infinite and thus $\sigma =\sigma _{sc}=\infty $, i.e., $\rho =0$ and the
normal state part of sample is short-circuited. On the other hand, the
resistivity in the TAFF regime is about one to three orders of
magnitude less than normal state\cite{Palstra1}$^{,}$\cite{Palstra2}. It
means that although the conductivity $\sigma _{sc}$ in this range is finite,
it is still much larger than $\sigma _{normal}$, and we can obtain $\sigma
\approx \sigma _{sc}$.

The Hall effect in the mixed state gives important insight in the flux flow.
In the following section, we will discuss the vortex dynamics of S-09 in
flux flow region. The field dependence of the longitudinal resistivity $\rho
_{xx}(H)$ for $H//c$ is shown in Fig. 4(a). Superconductivity is suppressed
by increasing magnetic field up to 35 T and the transition of $\rho _{xx}(H)$
are shifted to lower magnetic fields at higher temperature. At the lowest
measuring temperature ($T=0.3K$), normal state is recovered from
superconducting state when field is up to 30 T. Fig. 4(b) shows the Hall
resistivity at $T\leqslant 10K$. It can be seen that $\rho _{xy}(H)=0$ in
low field when temperature is below $T_{c}$. At higher field region close to
the superconducting transition, the absolute values of Hall resistivity
increase and gradually reach the $\rho _{xy}(H)$ curve obtained in the
normal state at temperatures slightly higher than $T_{c}$. The $\rho
_{xy}(H) $ in the mixed state shifts with increasing temperature to lower
fields. All of these features are similar to the $\rho _{xx}(H)$ results. On
the other hand, the normal state $\rho _{xy}(H)$ curves are very close to
linear, except for low field parts where there is slight nonlinearity with
negative curvatures that can be ascribed to skew scattering due to excess Fe.%
\cite{Lei HC2}

The sign of the Hall resistivity is positive in the mixed state as well as
in the normal state indicating hole type carriers. There is no sign reversal
for $\rho _{xy}(H)$ in the mixed state, which is a typical behavior for
hole- and electron-type cuprate superconductors below $T_{c}$.\cite{Hagen}$%
^{,}$\cite{Cagigal} Because Hall conductivity $\sigma _{xy}(H)$ [$=\rho
_{xy}(H)/(\rho _{xx}(H)^{2}+\rho _{xy}(H)^{2})\cong \rho _{xy}(H)/\rho
_{xx}(H)^{2}$, when $\rho _{xx}(H)\gg |\rho _{xy}(H)|$] is usually
insensitive to disorder by a general argument of the vortex dynamics, it is
convenient to discuss the Hall results using $\sigma _{xy}(H)$.\cite%
{Samoilov}$^{,}$\cite{Vinokur} There are two contributions to the $\sigma
_{xy}(H)$ in the mixed state:
\begin{equation}
\sigma _{xy}(H)=\sigma _{xy,n}(H)+\sigma _{xy,sc}(H)
\end{equation}%
where $\sigma _{xy,n}(H)$ is the conductivity of normal quasiparticles that
experience a Lorentz force inside and around the vortex core. This term has
the same sign as the normal state and is proportional to H. The second term $%
\sigma _{xy,sc}(H)$ is an anomalous contribution due to the motion of
vortices parallel to the electrical current density $\mathbf{j}$.

From the theory based on the time-dependent Ginzburg-Landau (TDGL) equation,
$\sigma _{xy,sc}(H)\varpropto 1/H$ and it could have a sign opposite to that
of $\sigma _{xy,n}(H)$.\cite{Dorsey}$^{,}$\cite{Kopnin} Furthermore, the $%
\sigma _{xy,sc}(H)$ is the dominant term at low field but at higher field $%
\sigma _{xy,n}(H)$ are important and could dominate over $\sigma _{xy,sc}(H)$%
. Therefore, if $\sigma _{xy,sc}(H)$ has a different sign when compared to $%
\sigma _{xy,n}(H)$, it is possible to observe a sign reversal in the Hall
effect in the superconducting state,\cite{Dorsey}$^{,}$\cite{Kopnin} as for
example in YBCO.\cite{Ginsberg} On the other hand, it can be easily seen in
Fig. 5(a) that the Hall conductivity decreases with increasing field and the
field dependence of $\sigma _{xy}(H)$ changes more rapidly than $1/H$. This
suggests that $\sigma _{xy}(H)$ is not independent of disorder in the strong
pinning regime.\cite{Matsuda} According to the theory proposed by Fukuyama,
Ebisawa, and Tsuzuki (FET),\cite{Fukuyama} the sign of $\sigma _{xy,sc}(H)$
is given by the sign of $sgn(e)\partial N(\mu )/\partial \mu |_{\mu =E_{F}}$%
, where $sgn(e)$ is the sign of the carrier, $N(\mu )$ is the density of
states, $\mu $ is the chemical potential, and $E_{F}$ is the Fermi energy.
On the other hand, in the phenomenological theory based on Ginsburg-Landau
equation and its gauge invariance,\cite{Aronov} the sign of the $\sigma
_{xy,sc}(H)$ is determined by the signs of $sgn(e)\partial \ln
T_{c}/\partial \mu $. In any case, the sign of the Hall effect in the mixed
state depends on the details of the band structure. For a complicated Fermi
surface, the signs of $\sigma _{xy,sc}(H)$ may be different from that of $%
\sigma _{xy,n}(H)$. Therefore, contrary to cuprate superconductors, the
difference of Fermi surface may be one origin of absence of sign reversal in
Fe(Te,S). On the other hand, cuprate superconductors are d-wave
superconductors, whereas, for Fe(Te,S), the gap function is unknown, but is
most likely s-wave. This difference could be another origin of different
contribution of the vortex cores to the Hall conductivity.\cite{Nagaoka}

\begin{figure}[tbp]
\centerline{\includegraphics[scale=0.95]{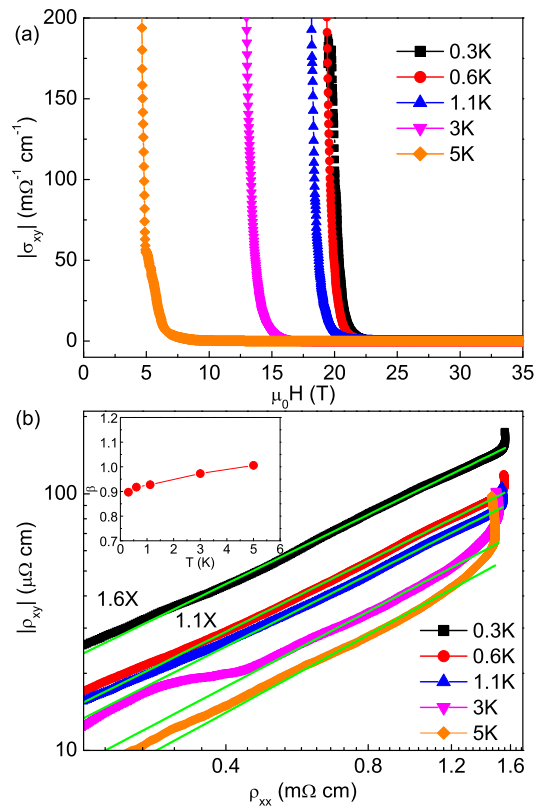}} \vspace*{-0.3cm}
\caption{(a) Field dependence of absolute values of the Hall conductivity $|%
\protect\sigma _{xy}(H)|$ measured at various temperatures in dc magnetic
fields up to 35T. (b) $|\protect\rho _{xy}|$ vs $\protect\rho _{xx}$ at
various temperatures. The solid lines are fitting results using the scaling
behavior $|\protect\rho _{xy}(H)|=A\protect\rho _{xx}(H)^{\protect\beta }$.
Inset of (b) shows the temperature dependence of $\protect\beta (T)$.}
\end{figure}

At high magnetic fields, we observe $|\rho _{xy}(H)|=A\rho _{xx}(H)^{\beta }$
scaling (Fig. 4(b)). The values of $\beta $ are in the range of $0.9-1.0$
and increase slightly with temperature. A phenomenological model considering
the effect of pinning on the Hall resistivity proposed by Vinokur et al.,%
\cite{Vinokur} gives the scaling index $\beta =2$ which, as well as Hall
conductivity, is independent on the degree of disorder. This is believed to
be a general feature of any vortex state with disorder dominated dynamics.
On the other hand, based on the normal core model proposed by Bardeen and
Stephen,\cite{Bardeen} Wang, Dong, and Ting (WDT),\cite{Wang ZD1}$^{,}$\cite%
{Wang ZD2} developed a theory for the Hall effect that includes both pinning
and thermal fluctuations. In the WDT theory the scaling behavior is
explained by taking into account the backflow current of vortices due to
pinning. Thereby $\beta $ changes from 2 to 1.5 as the pinning strength
increases.\cite{Wang ZD2} This has been observed in irradiated YBCO samples,
where $\beta $ was found to decrease from 1.5 compared to 2 after
irradiation,\cite{Kang} and in HgBa$_{2}$CaCu$_{2}$O$_{6+x}$ thin films with
columnar defects, where $\beta $ changes from 1.0 to 1.2 with increasing the
field.\cite{Kang2} Therefore, small values of $\beta $ in S-09 (inset of
Fig. 4(b)) may be connected with the strong pinning strength due to the
considerably large concentration of defects in Fe$_{1+y}$(Te$_{1-x}$S$_{x}$)$%
_{z}$.\cite{Hu RW}

\section{Conclusion}

In summary, we investigated the resistive TAFF and flux flow Hall effect of
Fe$_{1.14(1)}$(Te$_{0.91(2)}$S$_{0.09(2)}$)$_{z}$ single crystal in high and
stable magnetic fields up to 35 T. TAFF behavior could be understood within
the framework of modified Arrhenius relation assuming that the prefactor $%
\rho _{0f}$ is temperature-dependent while $\rho _{c}$ is
temperature-independent, and $U(T,H)=U_{0}(H)(1-t)^{q}$, $q$ can be set as a
free parameter for fitting not limited to 1. There is a crossover from
single-vortex pinning region to collective creep region for both field
direction. Furthermore, $q$ changes from 1 to 2 when magnetic field is
rotated along ab plane to c axis but it is not sensitive to the magnitude of
magnetic field below 35T. Hall and longitudinal resistivity in mixed state
indicate that there is no Hall sign change, as opposed to cuprate
superconductors. We observed scaling behavior $|\rho _{xy}(H)|=A\rho
_{xx}(H)^{\beta }$ with the scaling exponent $\beta $ about $1$, which may
be due to strong pinning strength in considerably disordered in Fe$_{1+y}$(Te%
$_{1-x}$S$_{x}$)$_{z}$ system.

\section{Acknowledgements}

We thank T. P. Murphy and J. B. Warren for experimental support at NHMFL and
Brookhaven National Laboratory (BNL). This work was carried out at BNL,
which is operated for the U.S. Department of Energy by Brookhaven Science
Associates DE-Ac02-98CH10886. This work was in part supported by the U.S.
Department of Energy, Office of Science, Office of Basic Energy Sciences as
part of the Energy Frontier Research Center (EFRC), Center for Emergent
Superconductivity (CES). A portion of this work was performed at NHMFL,
which is supported by NSF Cooperative Agreement No. DMR-0084173, by the
State of Florida, and by the U.S. Department of Energy.

$^{\ast }$Present address: Ames Laboratory US DOE and Department of Physics
and Astronomy, Iowa State University, Ames, IA 50011, USA.

\end{document}